\documentclass[twocolumn,showpacs,preprintnumbers,amsmath,amssymb,floatfix,prd,nofootinbib]{revtex4-1}
\usepackage{epsfig}
\usepackage{dcolumn}
\usepackage{bbold} 
\usepackage{wasysym} 
\usepackage{hyperref}
\usepackage{color,soul}
\def \be  {\begin{equation}}
\def \ee  {\end{equation}}
\def \ba  {\begin{eqnarray}}
\def \ea  {\end{eqnarray}}
\def \baa {\begin{eqnarray*}}
\def \eaa {\end{eqnarray*}}
\def \nn {\nonumber}

\newcommand\f{\frac}
\newcommand\jet{\text{jet}}

\begin{document}
\title{Using hadron-in-jet data in a global analysis of $D^{*}$ fragmentation functions}
%
%
\author{Daniele P.\ Anderle}
\email{daniele.anderle@manchester.ac.uk}
\affiliation{
Lancaster-Manchester-Sheffield Consortium for Fundamental Physics, School of Physics and
Astronomy, University of Manchester, Manchester, M13 9PL, U.K.}
\author{Tom Kaufmann}
\email{tom.kaufmann@uni-tuebingen.de}
\affiliation{Institute for Theoretical Physics, University of T\"ubingen, Auf der Morgenstelle 
14, 72076 T\"ubingen, Germany}
\author{Felix Ringer}
\email{fmringer@lbl.gov}
\affiliation{Nuclear Science Division, Lawrence Berkeley National Laboratory, Berkeley, CA 94720, USA}
\author{Marco Stratmann}
\email{marco.stratmann@uni-tuebingen.de}
\affiliation{Institute for Theoretical Physics, University of T\"ubingen, Auf der Morgenstelle 
14, 72076 T\"ubingen, Germany}

\author{Ivan Vitev}
\email{ivitev@lanl.gov}
\affiliation{Theoretical Division, Los Alamos National Laboratory, Los Alamos, New Mexico 87545, USA}

\begin{abstract}
We present a novel global QCD analysis of charged $D^{*}$-meson fragmentation functions at next-to-leading order
accuracy. This is achieved by making use of the available data for single-inclusive $D^{*}$-meson production in electron-positron
annihilation, hadron-hadron collisions, and, for the first time, in-jet fragmentation in proton-proton scattering.
It is shown how to include all relevant processes efficiently 
and without approximations within the Mellin moment technique, specifically for the in-jet fragmentation cross
section.
The presented technical framework is generic and can be 
straightforwardly applied to future analyses of fragmentation functions for
other hadron species, as soon as more in-jet fragmentation data become available.
We choose to work within the Zero Mass Variable Flavor Number Scheme which is applicable for sufficiently 
high energies and transverse momenta. 
The obtained optimum set of parton-to-$D^{*}$ fragmentation functions
is accompanied by Hessian uncertainty sets
which allow one to propagate hadronization uncertainties to other processes of interest.
\end{abstract}


\maketitle

\section{Introduction}
%
Cross sections at collider experiments can often be reliably calculated within the framework of 
perturbative Quantum Chromodynamics (pQCD). The crucial foundation for such computations
are so-called factorization theorems that allow for a systematic separation of perturbative 
and non-perturbative physics~\cite{ref:fact}. 
Well-known examples for the latter are parton distribution functions (PDFs)
that are, by now, rather tightly constrained by global QCD fits to data and are a
crucial asset in all scattering processes with hadrons in the initial-state.

Whenever an observable involves detected hadrons in the final-state, the theoretical calculation requires 
another type of non-perturbative functions as input. These parton-to-hadron fragmentation functions (FFs) 
describe the non-perturbative transition of a parton produced in the hard-scattering event 
into the observed hadron. Like PDFs, these functions were shown to be universal and can be 
only extracted from data through global QCD analyses. 
The knowledge of FFs for different hadron species and estimates of their uncertainties is therefore 
vital for precise theoretical calculations and, hence, has received quite some interest in the past; 
see, for instance, Ref.~\cite{Metz:2016swz} for a recent review. 

In this work, we consider the hadronization of quarks and gluons into heavy-flavored mesons,
more specifically, charged $D^*$-mesons, that are of particular relevance in the era of the LHC.
In general, the theoretical treatment of heavy quarks itself provides a unique laboratory to test pQCD. 
Correctly describing heavy flavor cross sections 
that have been measured both at very high energies at the LHC and at various low energy experiments 
poses unique challenges to our understanding of QCD. 
Charm production cross sections are used, for example, to constrain the gluon 
PDF at small-$x$~\cite{Gauld:2015yia}, and they play a vital role in cosmic-ray and 
neutrino astrophysics~\cite{ref:astro}.
Another important area of research concerns the modification of heavy flavor yields
in heavy-ion collisions~\cite{ref:heavyion} where highly energetic partons can traverse the quark-gluon plasma
thereby attaining valuable information about the properties of the QCD medium. 
For instance, the energy loss mechanisms, that allow for a quantitative description of in-medium effects, 
crucially depend on the underlying fragmentation process. 

In pQCD calculations, the heavy quark mass $m_Q$ introduces an additional large 
scale apart from some other hard scale characterizing the process,
such as a measured transverse momentum $p_T$. 
These multi-scale problems carry additional theoretical challenges as compared to 
processes involving only light quarks and gluons.
There are various approaches in the literature of how to deal with heavy quark masses 
in general and in the fragmentation process in particular.
In the context of $pp$ collisions relevant for LHC phenomenology the following schemes have been
put forward and used in their various kinematic regimes of applicability.
In the Fixed Flavor Number Scheme (FFNS)~\cite{ref:ffns}, the heavy quark $Q$ is not treated 
as an active parton in the proton but, instead, is solely produced extrinsically in the hard scattering.  
Logarithms of the ratio of the heavy quark mass $m_Q$ and the hard scale of the process, $p_T$, 
are only taken into account in fixed order perturbation theory. 
Therefore, this scheme is applicable in the region $p_T\sim m_Q$. 
The Zero Mass Variable Flavor Number Scheme (ZMVFNS), on the other hand, 
is only applicable in the limit $p_T\gg m_Q$. Here, large logarithms of $m_Q/p_T$ are resummed 
through DGLAP evolution equations to all orders. $m_Q$ is set to zero in all partonic cross sections, 
and the heavy quark is treated as an active, massless parton in the proton. 

In the context of fragmentation processes, the Fixed Order plus Next-to-Leading Logarithms prescription 
(FONLL)~\cite{ref:cacciari,Cacciari:2005uk} as well as the 
General Mass Variable Flavor Number Scheme (GMVFNS)~\cite{ref:gmvfns,Kneesch:2007ey} 
are examples of unified frameworks to cover both the high $p_T$ region, $p_T\gg m_Q$, 
and the low $p_T$ tail, $p_T\lesssim m_Q$, similar to the 
ZMVFNS and FFNS, respectively. In the FONLL approach, the FFs of heavy-flavored mesons
are separated into a perturbatively calculable parton-to-heavy quark $i\to Q$ contribution 
and a non-perturbative heavy quark-to-heavy meson $Q\to h$ piece that is fitted to data. 
This separation is possible as the heavy quark mass sets an additional scale in the perturbative regime. 
Instead, in the GMVFNS the entire parton-to-heavy meson FF is treated as a non-perturbative function 
and is extracted from the available data. 
We note that another scheme was developed recently 
in Ref.~\cite{Fickinger:2016rfd} within the framework of Soft Collinear Effective Theory (SCET). 

Since we are primarily interested in LHC phenomenology in this work, in particular
the impact of in-jet fragmentation data at $p_T\gg m_Q$, we choose to work in the ZMVFNS 
using purely non-perturbative FFs similar to the analyses of FFs for light hadron species.  
As will be discussed in detail below, the inclusive $p_T$-spectrum of charged $D^*$-mesons in $pp$ collisions 
can be fairly well described in the ZMVFNS down to rather low values of about $p_T\sim 5\,\mathrm{GeV}$
in spite of imposing a cut $p_T\ge 10\,\mathrm{GeV}$ when fitting $pp$ data.

Traditionally, the main reference process to determine FFs is semi-inclusive electron-positron annihilation (SIA),
$e^+e^-\to hX$. Here, $h$ denotes the detected hadron and $X$ the unobserved final-state remnant. 
To the best of our knowledge, all the approaches to heavy quark fragmentation mentioned above 
rely only on SIA data to determine the relevant non-perturbative input following similar non-global fits of
light hadron (pion, kaon) FFs \cite{ref:lightffs}.
While quark-to-hadron FFs can be relatively well constrained from SIA data, 
it is, in particular, the gluon-to-hadron FF that is at best only very poorly constrained by SIA data alone.
Therefore, global QCD analyses of light hadron FFs have also included vital proton-proton scattering data, 
$pp\to hX$, in order to better constrain the gluon FF. 
In addition, Semi-Inclusive Deep Inelastic Scattering (SIDIS), 
$\ell p\to \ell^\prime hX$, data are needed to perform a quark-antiquark and quark flavor separation of FFs.
Such global fits of light hadron FFs can be found in~\cite{ref:dss}.

In this paper, we will provide the first global QCD analysis of charmed-meson FFs following the
framework outlined by the DSS group in \cite{ref:dss} at next-to-leading order (NLO) accuracy
using the Mellin moment technique \cite{Stratmann:2001pb}.
We note that recently first efforts have been made to perform fits of light hadron FFs 
at next-to-next-to-leading order (NNLO) accuracy~\cite{Anderle:2015lqa,Bertone:2017tyb}, 
and also by including all-order resummations~\cite{Anderle:2012rq,Anderle:2016czy}. 
So far, these efforts have been limited to SIA data only due to the lack of other 
single-inclusive particle production cross sections at NNLO accuracy;
see, for example, \cite{Anderle:2016kwa} for the progress of an ongoing SIDIS calculation at NNLO.
As has become customary for both PDF and FF analyses these days, we also present an attempt to estimate
the remaining uncertainties of the extracted FF, 
for which we adopt the Hessian method \cite{ref:hessian}. The Hessian uncertainty sets can be used
to propagate hadronization uncertainties to any other processes of interest such as, for instance,
high-$p_T$ $D^*$-meson production in proton-nucleus collisions at the LHC or BNL-RHIC.

Besides the processes that are traditionally included in global analyses of FFs, 
like SIA and inclusive high-$p_T$ hadron production in $pp$ collisions, we also include 
for the first time in-jet fragmentation data from the LHC. 
Specifically, we include data for the ``jet fragmentation function'', $pp\to (\text{jet}\, h)X$, 
where a hadron is identified inside a jet. We consider the observable, where the longitudinal 
momentum distribution differential in $z_h=p_T/p_T^\jet$ is measured, 
with $p_T$ ($p_T^\jet$) denoting the hadron (jet) transverse momentum. 
The fact that at leading order (LO) accuracy the in-jet observable is directly
probing the $z=z_h$ dependence of FFs explains their potential relevance for analyses
of FFs. In-jet fragmentation was pioneered in~\cite{ref:injet} 
for exclusive jet samples. The extension to inclusive jet samples was developed 
in~\cite{Arleo:2013tya,Kaufmann:2015hma} within standard pQCD at NLO accuracy, 
allowing for a direct comparison with data from the LHC. 
In Ref.~\cite{Kang:2016ehg} the result was re-derived within SCET. 
Thanks to the effective field theory treatment, the additional all-order resummation 
of single logarithms in the jet-size parameter $\alpha_s^n\ln^n R$ was achieved, 
yielding consistent results at NLO+NLL$_{\text{R}}$ accuracy. 
In this work, we will work at NLO accuracy, and we leave a detailed study of the impact of 
NLO+NLL$_{\text{R}}$ corrections on fits of FFs for a dedicated, future publication. 

In Ref.~\cite{Chien:2015ctp}, it was found that the $D^{*}$-in-jet data from ATLAS~\cite{Aad:2011td} are
not well described by existing fits of $D^*$-meson FFs \cite{Kneesch:2007ey}
even though they give a good description of both SIA and inclusive $pp$ data;
see Ref.~\cite{Bain:2016clc} for related work. This leads to the important
question, which we address in detail in this work, if there is a real tension between the fitted data 
sets and the in-jet observable or if it is possible to accommodate all data sets in a combined, global fit. 
We note that apart from the $D^{*}$-in-jet data by ATLAS~\cite{Aad:2011td} there are also
in-jet results available from the LHC for unidentified light charged hadrons \cite{ref:unid-data},
mainly in heavy-ion collisions though, as well as for prompt and non-prompt $J/\psi$ production
in jets \cite{Aaij:2017fak}. In this first exploratory study of the impact of in-jet data
on fits of FFs we therefore limit ourselves to developing the necessary theoretical framework and to
a global analysis of parton-to-$D^{*+}$ and parton-to-$D^{*-}$ FFs utilizing the ATLAS in-jet data.
However, we wish to emphasize that the technical framework presented below
is generic and can be straightforwardly applied to future analyses of fragmentation functions for
other hadron species as soon as more in-jet fragmentation data become available.

Finally, we notice that various combined differential cross section data for charged $D^{*}$ mesons 
obtained in deep-inelastic lepton-proton collisions are available
from the H1 and ZEUS Collaborations \cite{H1:2015dma}. Since the data extend down to relatively
low values of transverse momentum and photon virtuality $Q$, they need to be described in
a theoretical framework which keeps the full dependence on the charm quark mass \cite{H1:2015dma}.
Hence, these data cannot be included in our current global QCD analysis that is based on 
the ZMVFNS approximation.

The remainder of this paper is organized as follows. 
In Section~\ref{sec:tech}, we discuss the technical framework for all three processes 
that are included in our global QCD analysis, namely $e^+e^-\to D^* X$, $pp\to D^* X$, 
and $pp\to (\text{jet}\, D^*)X$, with particular emphasis on the latter.
Note that throughout this paper, $D^{*}$ collectively denotes both charged
mesons, i.e.~$D^{*+}$ and/or $D^{*-}$. 
Next, in Section~\ref{sec:fit}, we briefly present the details of our analysis comprising 
the parametrization of the FFs at some input scale, the selection of experimental data and
cuts imposed on the fit, the Mellin moment technique used throughout this paper, 
and the Hessian uncertainty method. 
In Section~\ref{sec:pheno}, we present and discuss the results of our global analysis 
of parton-to-$D^{*}$ FFs at NLO accuracy, and compare the results of the fit to the available data. 
In addition, we compare to the previous fit provided by Ref.~\cite{Kneesch:2007ey}. 
In Section~\ref{sec:conclusions}, we draw our conclusions and present a brief outlook.
 
\section{Technical Framework \label{sec:tech}}

\subsection{Single-inclusive $e^+e^-$ annihilation \label{sec:SIA}}
%
The cross section for the single-inclusive annihilation process, $e^+e^-\to\gamma/Z\to h X$, 
is usually normalized to the total hadronic cross section $\sigma_{\text{tot}}$ 
and may be written schematically as 
\be\label{eq:TL}
\frac{1}{\sigma_\text{tot}} \frac{d\sigma^{e^+e^- \to hX}}{dz} = \frac{\sigma_0}{\sigma_{\text{tot}}}
\left[F_T^h(z,Q^2) + F_L^h(z,Q^2)\right]\,.
\ee
It is common to decompose the cross section (\ref{eq:TL}) into a transverse ($T$) and longitudinal ($L$) part
although this is of no practical relevance for $D$-meson production.
We have introduced the scaling variable
\be
\label{eq:zdef}
z\equiv \frac{2P_h\cdot q}{Q^2}\stackrel{\mathrm{c.m.s.}}{=} \frac{2E_h}{Q}\,, 
\ee
where $P_h$ and $q$ are the four momenta of the observed hadron and time-like $\gamma/Z$ boson, respectively.
Moreover, $Q^2 \equiv q^2 = S$. As is indicated in Eq.~(\ref{eq:zdef}), $z$ reduces to the hadron's energy fraction 
in the center-of-mass system (c.m.s.) frame and is often also labeled as $x_E$ \cite{ref:nason}. 
The total cross section for $e^+e^- \to $ hadrons at NLO accuracy reads
\be
\sigma_\text{tot} = \sum_q \hat{e}_q^2\, \sigma_0 \left[1 + \frac{\alpha_s(Q^2)}{\pi} \right]\,,
\ee
where $\alpha$ and $\alpha_s$ are the electromagnetic and the strong coupling, respectively, and $\sigma_0=4\pi \alpha^2(Q^2)/S$. We denote the electroweak quark charges by $\hat{e}_q^2$, which may be found, for instance, in 
App.\ A of Ref.~\cite{deFlorian:1997zj}.

To make factorization explicit, the transverse and longitudinal time-like structure functions 
in Eq.~\eqref{eq:TL} can be written as a convolution of perturbative coefficient functions 
$\mathbb{C}_i^k$, $i=q,\bar{q},g$~\cite{ref:sia-coefficients}, and non-perturbative FFs $D_i^h$,
\ba\label{eq:structurefunctions}
F_k^h(z,Q^2) &=& \sum_q \hat{e}_q^2 \Big\{ \left[\mathbb{C}_q^k \otimes (D_q^h + D_{\bar{q}}^h)\right](z,Q^2)\nn\\
&\phantom{=}& + \left[\mathbb{C}_g^k \otimes D_g^h\right](z,Q^2) \Big\} \,,
\ea
where $k=T,L$. The standard convolution integral with respect to the first argument
is denoted by the symbol $\otimes$ and reads
\be\label{eq:convolution}
[f\otimes g] (z,\dots)\equiv \int_0^1dx\int_0^1dy \, f(x,\dots)\, g(y,\dots)\,\delta(z-xy)
\ee
for two arbitrary functions $f$ and $g$.
%

As always, the notion of factorization as applied in Eq.~\eqref{eq:structurefunctions} 
is only valid up to corrections proportional to inverse powers of the hard scale \cite{ref:nason}.
For a one-scale process like SIA, the hard scale should be chosen to be of
${\cal{O}}(Q)$ and $Q$ itself should be at least of ${\cal{O}}$(few GeV).
For simplicity, we have chosen the factorization 
and renormalization scales in Eq.~\eqref{eq:structurefunctions} 
equal to the hard scale, i.e., $\mu_R=\mu_F\equiv Q$.

Kinematical effects related to the non-zero mass $m_h$ of the produced hadron $h$ are
another source of corrections to the factorized framework where $m_h$ is neglected throughout.
Deviations of the data from theory are expected to show up at the lower end of the $z$-spectrum, as we
shall see in the phenomenological section, and are more pronounced for heavier than for light mesons.
One usually introduces a cut $z_{\min}$ in global analyses of FFs \cite{ref:lightffs,ref:dss}
below which the data cannot be used and the theory outlined above is not valid.
Such a cut also avoids the region in $z$ where fixed-order evolution kernels receive large
logarithmic corrections which otherwise can only be dealt with by all-order resummations,
see, for instance, Ref.~\cite{Anderle:2016czy}.

\subsection{Single-inclusive $D^{*}$ production $pp$ collisions \label{sec:pphX}}
%
The production of high-$p_T$ hadrons in hadronic collisions offers valuable
and complementary information compared to SIA data in global QCD analyses of FFs. 
The dominance of the $gg\to gX$ and $qg\to gX$ partonic subprocesses at not too large
values of $p_T$ gives access to the gluon-to-hadron fragmentation function, which is
only very indirectly accessible in SIA through scaling violations and, hence, largely
unconstrained. 

In addition to data for the sum of charged $D^*$ mesons, $D^{*+} + D^{*-}\equiv D^{*\pm}$,
from ATLAS \cite{Aad:2015zix}
and LHCb \cite{Aaij:2013mga,Aaij:2015bpa}, measurements of positively charged $D^{*+}$ mesons
are available from both the ALICE \cite{ALICE:2011aa, Abelev:2012vra} and 
CDF \cite{Acosta:2003ax} collaborations. 
The latter sets of data offer new information on the charge separation of $D^*$ meson FFs
that is not available from SIA where only the sum $D^{*\pm}$ can be observed. 
It is also worth recalling that the Tevatron data from CDF \cite{Acosta:2003ax} are taken in 
$p\bar{p}$ rather than $pp$ collisions and that LHCb has the unique capability to perform
measurements at different asymmetric, forward rapidity intervals \cite{Aaij:2013mga,Aaij:2015bpa}.
Both sets of data will add unique information to our global analysis.

The factorized cross section for a given hadron $p_T$ and pseudorapidity $\eta$ 
may schematically be written as a convolution of appropriately combined
PDFs, parton-to-hadron FFs, and partonic hard scattering cross sections:
\ba\label{eq:xsec_pphx}
\frac{d\sigma^{H_1 H_2\to h X}}{dp_Td\eta}&=&\frac{2p_T}{S}
\sum_{abc} f_a^{H_1} \otimes f_b^{H_2} \otimes d\hat{\sigma}_{ab}^c \otimes D_c^h\,.
\ea
Here, $f_a^{H_1}$ and $f_b^{H_2}$ denote the PDFs with flavor $a$ and $b$ in hadron $H_1$ and $H_2$,
respectively, and $D_c^h$ is the $c\to h$ FF.
The sum in (\ref{eq:xsec_pphx}) is over all contributing partonic cross sections $a b \to c X$,
denoted as $d\hat{\sigma}_{ab}^c$, which may be calculated as a perturbative series in $\alpha_s$, 
starting at $\mathcal{O}(\alpha_s^2)$ which corresponds to the LO approximation. 
Hence, to perform a consistent NLO analysis of fragmentation functions, we include the
$\mathcal{O}(\alpha_s^3)$ corrections which have been computed analytically 
in~\cite{ref:ppxsec-nlo}.
As mentioned above, the factorized form given in Eq.~\eqref{eq:xsec_pphx} is again only valid up 
to power corrections that are suppressed by inverse powers of the hard scale, in this case $p_T$. 
Throughout this work, we choose the factorization and renormalization scales for this process
to be equal to the transverse momentum of the observed hadron, i.e., $\mu = p_T$, but we will
illustrate the residual dependence on $\mu$ in the phenomenological section below by varying
$\mu$ by the conventional factor of two up and down.

One drawback of the single-inclusive high-$p_T$ production process is 
that the information on the $z$ dependence of the probed FFs is only accessible 
in integrated form through one of the convolution integrals in Eq.~\eqref{eq:xsec_pphx}.
The range of integration allowed by kinematical considerations for a given $p_T$ and $\eta$ 
of the observed hadron is rather broad and may reach well below the cut $z_{\min}$ mentioned above. 
However, it has been shown in Ref.~\cite{Sassot:2010bh} that one 
samples on average predominantly fairly large values of $z$ in Eq.~\eqref{eq:xsec_pphx}, 
$\langle z \rangle \simeq 0.4$ at mid rapidity and further increasing towards forward rapidities, 
and that $z$ values below $z_{\min}$ are irrelevant for all practical purposes.
Considering hadrons inside jets
rather than single-inclusive hadron production allows one to sample $z$ more directly,
as we shall discuss in some detail next.

\subsection{$D^{*}$ meson in-jet production \label{sec:jethX}}
%
The inclusive production of identified hadrons inside a fully reconstructed jet $pp\to (\jet\, h)X$, 
where the hadron is part of the jet, has been studied for $pp$ collisions 
in Refs.~\cite{Arleo:2013tya,Kaufmann:2015hma,Kang:2016ehg}. 
In~\cite{Arleo:2013tya}, the NLO cross section was obtained using a Monte-Carlo (MC) phase space integrator. 
Instead, in Refs.~\cite{Kaufmann:2015hma,Kang:2016ehg} analytical results were obtained 
using the approximation that the jet is sufficiently collimated. 
The NLO result of~\cite{Kaufmann:2015hma} was derived within the standard pQCD framework, 
whereas~\cite{Kang:2016ehg} employed methods within SCET for inclusive jet production~\cite{Kang:2016mcy,Dai:2016hz}, 
which allows for the additional resummation of single logarithms of the jet size parameter $R$. 
At NLO accuracy, the analytical result of the cross section can be 
schematically written as $\mathcal{A} + \mathcal{B} \log R + \mathcal{O}(R^2)$. 
If the jet is sufficiently narrow, i.e., $R\ll 1$, power corrections of the order ${\cal O}(R^2)$ can be neglected. 
In studies for inclusive jet production~\cite{ref:inclJets}, 
it was found that this ``narrow jet approximation'' is valid even for relatively large values of $R$. 
For example, for $R=0.7$ the agreement between the thus obtained analytical results and the full MC result at NLO 
is better than 5\%. This observation was also confirmed for the
in-jet production of hadrons in Ref.~\cite{Kaufmann:2015hma} by comparing to the 
full NLO MC calculation of~\cite{Arleo:2013tya}. 

In this work, we need the hadron-in-jet results for the anti-$k_T$ jet algorithm \cite{Cacciari:2008gp}. 
Currently, the only available data set for $D^{*\pm}$ mesons within jets is provided by the 
ATLAS collaboration~\cite{Aad:2011td} for which the anti-$k_T$ algorithm was used with a jet size parameter of $R=0.6$. 
However, the results for cone \cite{Salam:2007xv} and $J_{E_T}$ \cite{Georgi:2014zwa} jets 
are also available in the literature~\cite{Kaufmann:2015hma,Kang:2016ehg,Kang:2017mda}. 

As it was discussed in detail in Ref.~\cite{Kaufmann:2015hma}, the in-jet fragmentation provides a more direct access to the  $z$-dependence of FFs than
data on single-inclusive hadron production. At LO accuracy, 
the cross section is directly proportional to the FFs probed at the momentum fraction $z=z_h$, where 
\be
z_h \equiv \frac{p_T}{p_T^\jet}
\ee
and $p_T$ ($p_T^\jet$) denotes the transverse momentum of the hadron (jet).   
The cross section for the process $pp\to (\jet\, h)X$ may be written as
\ba\label{eq:xsec_ppjethX}
\frac{d\sigma^{pp\rightarrow (\jet\,h)X}}{dp_T^{\jet} d\eta^\jet d z_h} &=&
\frac{2p_T^{\jet}}{S} \sum_{a,b,c} \int_{x_a^{\mathrm{min}}}^1 \frac{dx_a}{x_a} f_a(x_a,\mu)\nn\\[2mm]
&\phantom{=}&\hspace*{-2.5cm}\times\, \int_{x_b^{\mathrm{min}}}^1 \frac{dx_b}{x_b}  f_b(x_b,\mu) \int_{z_c^{\mathrm{min}}}^1 \frac{dz_c}{z_c^2} \, \frac{d\hat{\sigma}_{ab}^c(\hat{s},\hat{p}_T,\hat\eta,\mu)}{vdvdw}\nn\\[2mm]
&\phantom{=}&\hspace*{-2.5cm}\times\, \mathcal{G}_c^h(z_c,z_h,\mu,R)\,,
\ea
where, again, we have set the renormalization and factorization scales to be equal 
and collectively denoted them by $\mu$. For this process, we choose $\mu=p_T^\jet$ as
our default choice of scale. The partonic 
cross sections $d\hat{\sigma}_{ab}^c$ are the same as they appear in the cross section
for single-inclusive hadron production in Eq.~\eqref{eq:xsec_pphx}. These hard functions
depend on the jet partonic transverse momentum $\hat{p}_T = p_T^\jet / z_c$, the partonic rapidity 
$\hat{\eta} = \eta^\jet - \log(x_a / x_b)/2$ and the partonic c.m.s.\ energy squared $\hat{s} = x_a x_b S$
with $\sqrt{S}$ the hadronic c.m.s.\ energy.
The integration limits are customarily expressed in terms of the hadronic variables
\be\label{eq:VW}
V \equiv 1 - \frac{p_T^\jet}{\sqrt{S}}\, e^{-\eta^\jet} \,,\quad W \equiv \frac{(p_T^\jet)^2}{SV(1-V)}\,,
\ee
and read
\ba\label{eq:xaxbzc}
x_a^{\mathrm{min}}&=&W\,,\quad x_b^{\mathrm{min}}\,=\,\frac{1-V}{1-VW/x_a}\,,\nn\\
z_c^{\mathrm{min}}&=&\frac{1-V}{x_b}+\frac{VW}{x_a}\,.
\ea
The function $\mathcal{G}_c^h$ in Eq.~\eqref{eq:xsec_pphx}
contains all the information on the production of the final-state jet and the identified hadron inside the jet and, hence,
depends on the jet size parameter $R$. To NLO accuracy, we can further decompose ${\cal G}_c^h$ as
\ba\label{eq:G_QCD}
\mathcal{G}_c^{h}(z_c,z_h,\mu,R) &=& \sum_e j_{c\to e}(z_c,R,\mu)
\nn\\[2mm]
&\phantom{=}& \hspace*{-2cm}  \times\sum_{c'}\int_{z_h}^1 \frac{d\xi}{\xi}\,
\tilde{j}_{e\to c'}(\xi,R,\mu )\, D_{c'}^h\left(\frac{z_h}{\xi},\mu \right).
\ea
The jet functions $j$ and $\tilde{j}$ describe the formation of the jet and the partonic fragmentation, respectively, and
may be found in Ref.~\cite{Kaufmann:2015hma}.
Inserting Eq.~\eqref{eq:G_QCD} into Eq.~\eqref{eq:xsec_ppjethX}, we may write the cross section as
\be\label{eq:G_QCD2}
\frac{d\sigma^{pp\rightarrow (\jet\,h)X}}{dp_T^{\jet} d\eta^\jet d z_h} = \sum_{e,c^\prime} \mathcal{E}_e 
\times \left[\tilde{j}_{e\to c^\prime} \otimes D_{c^\prime}^h \right](z_h)
\ee
where $\mathcal{E}_e$ contains all the sums and integrals over the PDFs, 
the partonic cross sections and the jet functions $j_{c\to e}$ and may be regarded as an ``effective charge" weighting
the different channels. The fragmentation functions appear in an actual convolution with the jet functions
$\tilde{j}$ with respect to $z_h$, multiplied by these effective charges. 
Eq.~\eqref{eq:G_QCD2} illustrates the structural similarity of the in-jet fragmentation cross section 
and SIA, enabling access to the $z$-dependence of the FFs.
Due to the hadronic initial-state, the gluon fragmentation function already appears at LO accuracy, as it is the
case for single-inclusive hadron production in $pp$ collisions.

Typically, the hadron-in-jet production data are normalized to the 
inclusive jet cross section $pp\to\text{jet}X$. Hence, the actual experimental observable is given by
\be
\label{eq:injet-obs}
F(z_h,p_T^\jet,\eta^\jet) \equiv \frac{d\sigma^{pp\rightarrow (\jet\,h)X}}{dp_T^{\jet} d\eta^\jet d z_h} 
\Big/ \frac{d\sigma^{pp\to \jet\, X}}{dp_T^\jet d\eta^\jet}\,.
\ee
It was found in \cite{Kaufmann:2015hma,Kang:2016mcy} that the cross section for inclusive jet
production may be written in a similar form as the single-inclusive hadron production
cross section in Eq.~\eqref{eq:xsec_pphx}, with only the fragmentation functions $D_c^h$ 
replaced by perturbatively calculable jet functions $J_c$, i.e.,
\be\label{eq:xsec_ppjetX}
\frac{d\sigma^{H_1 H_2\to \jet\, X}}{dp_T^\jet d\eta^\jet} = \frac{2p_T^\jet}{S}
\sum_{abc} f_a^{H_1} \otimes f_b^{H_2} \otimes d\hat{\sigma}_{ab}^c \otimes J_c\,.
\ee
Thus, we may use the numerically efficient codes of Refs.~\cite{ref:inclJets} 
to compute the hadron-in-jet cross section observable (\ref{eq:injet-obs}) in our global analysis of
$D^{*\pm}$ FFs.
%

\section{Outline of the analysis \label{sec:fit}}
%
\subsection{Parametrization \label{subsec:parametrization}}
%
As we choose to work in the ZMVFNS for our global analysis of $D^{*}$ FFs,
we closely follow the procedures for light hadron (pion and kaon) FFs as
outlined in Refs.~\cite{ref:dss,Anderle:2015lqa}.
However, due to the significantly smaller amount of data for $D^{*}$ production,
we adopt a slightly less flexible, more economical functional form to parametrize
the non-perturbative parton-to-$D^{*+}$ FFs at some initial scale $\mu_0$ 
in the commonly adopted $\overline{\mathrm{MS}}$ scheme:
\be\label{eq:Dparam}
D_i^{D^{*+}}(z,\mu_0^2)=\f{N_i\, z^{\alpha_i}(1-z)^{\beta_i}}
{B[2+\alpha_i,\beta_i+1]} \,.
\ee
We have tested that Eq.~\eqref{eq:Dparam} nevertheless yields a very satisfactory description of the data,
see also our results in Sec.~\ref{sec:pheno} below. The much simpler functional form with significantly less
parameters also has the additional benefit of greatly facilitating the fitting procedure and
the determination of uncertainties with the Hessian method.

We choose our initial scale to be equal to the charm quark mass $\mu_0 = m_c$. 
As we adopt the CT14 set of NLO PDFs and determination of the strong coupling $\alpha_s$~\cite{Dulat:2015mca}
in all our calculations of hadronic cross sections and the scale evolution of FFs, 
we also use the heavy quark masses according to CT14, i.e., $m_c = 1.3\,\mathrm{GeV}$ and $m_b = 4.75\,\mathrm{GeV}$.
Furthermore, we assume that at the initial scale $\mu_0$ the FFs for
all light quarks and antiquarks as well as for the anti-charm quark vanish, i.e.,
\be
D_q^{D^{*+}}(z,\mu_0^2) = 0,~~~~~~\text{ for } q=u,\bar{u},d,\bar{d},s,\bar{s},\bar{c}\,,
\ee
which has no impact on the quality of the fit. In any case, none of these FFs can be reliably determined
from the existing sets of data.

The bottom quark and antiquark FFs are included in the scale evolution above $\mu=m_b$, and, as
non-perturbative input, we only parametrize the total bottom-to-$D^{*+}$ fragmentation function 
$D_{b_\text{tot}}^{D^{*+}}(z,m_b^2) \equiv D_{b+\bar{b}}^{D^{*+}}(z,m_b^2)$. 
In total this leaves us with 9 non-zero parameters in Eq.~\eqref{eq:Dparam}
for $i=g,c,b_\text{tot}$ which is further reduced to 8 actual parameters to be determined in our
global analysis since it turns out
that $\beta_g$ is essentially unconstrained by data and has to be fixed.

As usual, the FFs for positively and negatively charged mesons are assumed to be related by charge conjugation, i.e.,
\be
D_q^{D^{*-}}(z,\mu^2) = D_{\bar{q}}^{D^{*+}}(z,\mu^2) 
\ee
for quarks and
\be
D_g^{D^{*-}}(z,\mu^2) = D_g^{D^{*+}}(z,\mu^2) 
\ee
for the gluon. This will be used to compute cross sections for all data sets which 
observe only the sum of charges $D^{*\pm}$.

Finally, the parametrization in Eq.~\eqref{eq:Dparam} is normalized to the respective 
$N=2$ Mellin moment by the denominator containing the Euler Beta function $B[a,b]$.
Hence, the coefficients $N_i$ constitute the contribution of $z D_i^{D^{*+}}$ 
to the energy-momentum sum rule
\be
\sum_h \int_0^1 dz\, z D_i^{h}(z,\mu^2) = 1\,.
\ee

\subsection{Selection of data sets \label{subsec:data}}
%
Numerous experimental data exist for the three types of processes described in Sec.~\ref{sec:tech}.
Identified $D^{*\pm}$ mesons in $e^+e^-$ collisions have been measured
both by the ALEPH \cite{Barate:1999bg} and OPAL \cite{Akers:1994jc, Ackerstaff:1997ki} collaborations at LEP at
a c.m.s.\ energy of $Q=M_Z$, the mass of the $Z$ boson.
Unfortunately, the results from the more recent OPAL analysis~\cite{Ackerstaff:1997ki}
are presented only in graphical form, and the corresponding numerical values are not anymore
available \cite{ref:OPAL_pc}. Thus, we decide to use only the older set of OPAL data~\cite{Akers:1994jc}
with less statistics and somewhat larger uncertainties in our fit.
Both collaborations also present bottom and charm flavor
tagged data. Here, only the OPAL collaboration provides numerical values which we include in our global analysis.

Both data sets from ALEPH and OPAL are not corrected for the branching ratios of the decay channels
used for the identification of the $D^{*\pm}$ mesons. 
To obtain properly normalized cross sections, we divide the data by the branching ratios $B_1$ and $B_2$ 
which are given by \cite{Olive:2016xmw}
\ba
B_1(D^{*+} \to D^0 \pi^+) &=& (67.7 \pm 0.5)\%\,,\nn\\
B_2(D^{0} \to K^- \pi^+) &=& (3.93 \pm 0.04)\%\,.
\ea
The uncertainties of the branching rations, $\Delta B_1$ and $\Delta B_2$, are propagated
into the systematic uncertainty of the SIA cross section data by adding them in quadrature, i.e.,
\be
\Delta d\sigma^\text{sys} = \sqrt{\left(\frac{d\sigma \Delta B_1}{B_1^2 B_2}\right)^2 
+ \left(\frac{d\sigma \Delta B_2}{B_1 B_2^2}\right)^2
+ \left(\frac{\Delta d\sigma}{B_1 B_2}\right)^2} \,.
\ee
Here, $\Delta d\sigma$ denotes the systematic error of the SIA data as provided by the ALEPH and OPAL experiments.

At lower c.m.s.\ energies, there are several measurements available around 
$Q \approx 30$ GeV~\cite{ref:other-sia}.
However, these data sets are rather old, and they consist of only a few data points that 
have very large uncertainties. Therefore, these sets do not add any relevant additional constraints 
to our global analysis, and, for simplicity, we choose to not include them.

Finally, some $e^+e^-$ experiments have measured $D^{*\pm}$ production just below the 
bottom threshold at around $Q\approx10.5$ GeV.
The most recent and precise data are from the BELLE collaboration \cite{Seuster:2005tr}.
However, as stated on the {\sc HepData} webpage, the ``data for this record have been removed at the request
of the authors due to an unrecoverable error in the measurement"; see \cite{ref:BELLEremoved}.
Hence, we have to refrain from using this data set in our fit, which, potentially, could have been a very
promising constraint from SIA in addition to the LEP data at $Q=M_Z$. We note that the previous
analysis of $D$-meson FFs by the KKKS group~\cite{Kneesch:2007ey}, to which we compare later on,
includes the BELLE data as they were not yet withdrawn at the time when their fit was performed.
Furthermore, CLEO \cite{Artuso:2004pj} and ARGUS \cite{Albrecht:1991ss}
also provide data measured at similar c.m.s.\ energies as BELLE. 
However, both data sets are not corrected for initial-state 
radiation (ISR) effects. In addition, the ARGUS data points have very large uncertainties.
The CLEO data have been included in the extractions of $D^*$ FFs in Refs.~\cite{Cacciari:2005uk,Kneesch:2007ey}.
While Ref.~\cite{Cacciari:2005uk} models the ISR effects based on data, the extraction of~\cite{Kneesch:2007ey} 
includes ISR using certain approximations in the theory calculation of the cross section. However,
both find noticeable tensions between the CLEO and ALEPH data.
Since this may or may not be related to the treatment of ISR corrections, we
choose not to include any of the low-energy SIA data in our analysis.

Data for inclusive $D^{*\pm}$ production in hadronic collisions are available from the CDF collaboration 
at the Tevatron \cite{Acosta:2003ax} and from the ALICE \cite{ALICE:2011aa, Abelev:2012vra},
ATLAS \cite{Aad:2015zix}, and LHCb \cite{Aaij:2013mga,Aaij:2015bpa} collaborations at the LHC.
We utilize all of these data sets in our global QCD analysis, as they provide valuable constraints on the
gluon fragmentation function. As was mentioned in the Introduction and in
Sec.~\ref{sec:jethX}, an important new asset of our analysis are the in-jet fragmentation data
for which ATLAS has presented results for identified $D^{*\pm}$ mesons inside
fully reconstructed jets~\cite{Aad:2011td}.

To ensure the validity of the ZMVFNS approximation and the massless treatment of
the $D^{*}$ mesons in the factorized formalism used to describe fragmentation processes,
we have to impose certain cuts on the above mentioned data sets.
For SIA, we only use data in the interval $0.1 < z < 0.95$, i.e., $z_{\min}=0.1$,
which is sufficient for the LEP data taken at $Q=M_Z$.
For all $p_T$-spectra of $D^{*}$ mesons in hadronic collisions we select a 
very conservative cut of $p_T>p_T^{\min}=10\,\mathrm{GeV}$ below which we exclude all data from the fit.
Notice that this cut forces us to exclude the LHCb data sets from the 7 TeV run; we nevertheless show
a comparison of our optimum fit to these data in Sec.~\ref{sec:pheno}.
We are confident that our resulting set of FFs is not affected by our choice of
$p_T^{\min}$ since we find that lowering the cut down to $5\,\mathrm{GeV}$ does not lead
to any significant changes in both the quality of the fit and the obtained optimum fit parameters in Eq.~\eqref{eq:Dparam}.
This also implies that our results can be reliably extrapolated down to $p_T$ values of about 
$5\,\mathrm{GeV}$, as we will also illustrate in some detail in Sec.~\ref{sec:pheno}.

\subsection{Mellin Moment Technique \label{subsec:mellin}}
%
As mentioned above, we work entirely in complex Mellin $N$ moment space in order to solve the scale evolution equations 
of the FFs, to compute the relevant SIA and $pp$ cross sections discussed in Sec.~\ref{sec:tech}, and to perform the 
actual fit and error analysis. 
The Mellin integral transform is well suited for these tasks as convolution integrals turn in ordinary products
in Mellin $N$ space and the integro-differential evolution equations can be solved analytically. 
The resulting numerical codes for global QCD analyses are very efficient and fast.

In general, the pair of Mellin integral and inverse transforms of a function $f(z)$ and $f(N)$ are defined by
\begin{equation}
\label{eq:mellin}
f(N)=\int_0^1dz\,z^{N-1}f(z) 
\end{equation}
and
\be\label{eq:inverse}
f(z) =\f{1}{2\pi i}\int_{{\cal C}_N} dN\, z^{-N}\, f(N)\,,
\ee
respectively, where ${\cal C}_N$ denotes a suitable contour in the complex Mellin $N$ plane that guarantees
fast convergence, see Refs.~\cite{ref:pegasus,Anderle:2016czy,Anderle:2016kwa} for a comprehensive discussion of
technical details and subtleties.
In practice, one ends up having to compute only a limited number of moments 
along the contour ${\cal C}_N$ in order to numerically solve the integral in Eq.~\eqref{eq:inverse}.

Our analysis is set up in the following way: for each data point we use the analytical ``truncated'' solution 
of the evolution equations at NLO accuracy in Mellin space, see, e.g., Refs.~\cite{ref:pegasus,Anderle:2016czy,Anderle:2016kwa},
to evolve the input FFs in \eqref{eq:Dparam} to the relevant scale. Next, the FFs are
combined with appropriate $N$ space expressions for the hard scattering subprocesses before the inverse
transform in Eq.~\eqref{eq:inverse} is performed numerically to evaluate the quality of the fit, see the next subsection.
More specifically, in case of SIA, see Sec.~\ref{sec:SIA}, this is achieved by taking the Mellin moments
of Eq.~\eqref{eq:structurefunctions} analytically, and all convolutions of FFs and coefficient functions
turn schematically into 
\be\label{eq:inverse1}
D(z) \otimes \mathbb{C}(z)=\f{1}{2\pi i}\int_{{\cal C}_N} dN\, z^{-N}\, D(N)\,\mathbb{C}\left(N\right) \, .
\ee
Here, each coefficient function $\mathbb{C}(N)$ can be evaluated explicitly using the general definition in Eq.~\eqref{eq:mellin} 
and appropriate analytic continuations of harmonic sums to non-integer, complex $N$ values, 
see, for instance, Ref.~\cite{Anderle:2016czy}.

For the more complicated expressions in $pp$ scattering discussed in Secs.~\ref{sec:pphX} and~\ref{sec:jethX}, 
one has to invoke an intermediate step as it is no longer possible or too cumbersome to perform the Mellin transform 
of the hard scattering cross sections analytically.
Instead, we follow the steps outlined in Ref.~\cite{Stratmann:2001pb,ref:dss} and first express the 
FFs that appear, e.g., in Eq.~\eqref{eq:xsec_pphx} in terms of their respective Mellin inverse, see Eq.~\eqref{eq:inverse}.
After some reordering, the inclusive hadron production cross section $H_1 H_2\to hX$ can be recasted as follows
\ba\label{eq:ppgrid}
\frac{d\sigma^{H_1 H_2\to h X}}{dp_Td\eta}&=&\sum_c \frac{1}{2\pi i} \int_{\mathcal{C}_N}dN\,D_c^h(N)\\ \nn 
&\phantom{=}&\times \frac{2p_T}{S}
\sum_{ab} f_a^{H_1} \otimes f_b^{H_2} \otimes d\hat{\sigma}_{ab}^c \otimes \tilde{D}_c^h\,,
\ea
where $\tilde{D}_c^h(z) = z^{-N}$. The second line is independent of the FFs we are interested in, only needs to
be evaluated once, and can be stored on a grid. In the end, for each data point one only has to perform
the remaining contour integral in \eqref{eq:ppgrid}. This method is completely general and does not
require any approximations such as $K$-factors, and is also employed
for the in-jet fragmentation $pp\to(\text{jet}\, h)X$ in Sec.~\ref{sec:jethX}.

\subsection{Fitting and the Hessian Uncertainty Method \label{subsec:errors}}
%
We obtain the optimum values for the eight free fit parameters in Eq.~\eqref{eq:Dparam}
by a standard $\chi^2$ minimization.
We define the $\chi^2$ for the $M$ data sets included in the fit, 
each containing $M_i$ data points that pass the selection cuts specified in
Sec.~\ref{subsec:data}, to be
\be
\label{eq:chi2}
\chi^2 = \sum_{i=1}^M \left[\frac{(1-\mathcal{N}_i)^2}{\Delta \mathcal{N}_i^2} 
+ \sum_{j=1}^{M_i}\frac{(\mathcal{N}_i T_j - E_j)^2}{\Delta E_j^2}\right]\;,
\ee
where $E_j$ is the experimental value for a given observable with uncertainty $\Delta E_j$
and $T_j$ is the corresponding theory calculation.
Furthermore, we have introduced normalization shifts $\mathcal{N}_i$ to account for this type of 
uncertainty whenever the normalization error $\Delta \mathcal{N}_i$ is stated by the experiments. 
The optimum normalization shifts $\mathcal{N}_i$ are computed analytically from the condition
that they should minimize the $\chi^2$. We note that we combine systematical and statistical
uncertainties in quadrature in $\Delta E_j$.

In order to estimate the uncertainties of the extracted FFs due to the experimental
uncertainties $\Delta E_j$ and $\Delta \mathcal{N}_i$, 
we adopt the widely used iterative Hessian approach \cite{ref:hessian} to explore
the range of possible variations of the obtained optimum parameters in the vicinity of the
minimum of the $\chi^2$ function for a given tolerance $\Delta \chi^2$. 
To this end, we provide 16 eigenvector sets for our FFs that correspond to the $+$ and $-$ directions 
of the eigenvectors of the diagonalized Hessian matrix.  
These sets greatly facilitate the propagation of hadronization uncertainties to
any observable of interest.
In fact, the uncertainty of an observable $\mathcal{O}$ may be calculated 
straightforwardly as \cite{ref:hessian}
\be
\Delta \mathcal{O} = \frac{1}{2}  \sqrt{ \sum_{i=1}^{8} \left[\mathcal{O}_{+i} - \mathcal{O}_{-i}\right]^2}\,,
\ee
where $\mathcal{O}_{\pm i}$ denote the observable calculated with
the plus or minus Hessian eigenvector set $i$, respectively.

Finally, we note that choosing the tolerance $\Delta \chi^2$ is to some extent arbitrary
in the presence of non-Gaussian or unaccounted uncertainties accompanying any global fit
of PDFs or FFs. We have made sure that our Hessian sets, computed with $\Delta\chi^2 = 4$,
faithfully reflect the experimental uncertainties of the SIA data, as can be seen
and will be discussed in the phenomenological section below. In this sense they correspond to 
uncertainties at the $68\%$ confidence level in the $z$ range that is constrained by data.
Outside that range, the uncertainties are biased by the choice and flexibility of the selected
functional form and assumptions made on the parameter space. 
In what follows, we will also briefly discuss additional, theoretical sources of 
uncertainties such as the choice and uncertainties of PDFs and from 
variations of the renormalization and factorization scales $\mu$. 

\section{Results \label{sec:pheno}}
\subsection{Parton-to-$D^{*+}$ fragmentation functions}

In this section, we present the results of our global determination of the parton-to-$D^{*+}$-meson
fragmentation functions and compare them to the previous fit of SIA data provided by Ref~\cite{Kneesch:2007ey},
which will be labeled ``KKKS08". Note, that we use the public available numerical code from~\cite{ref:KKKS08download}. 
In Tab.~\ref{tab:fitpar}, we list the numerical values of the parameters of our optimal fit at NLO accuracy, 
see Eq.~\eqref{eq:Dparam}. 
As already mentioned, the parameter $\beta_g$, which controls the $z\to 1$ behavior of
the gluon FF, is basically unconstrained by data. 
For this reason, we decided to keep $\beta_g=10$ fixed. Note that other choices, 
like $\beta_g=5$ or 15 yield a total $\chi^2$ which differs by less than one unit, which is well within
our tolerance $\Delta \chi^2=4$. It is worth mentioning that in all fits with different values of $\beta_g$, the
parameter $\alpha_g$ changes in such a way that the normalization $N_g$ remains essentially the same. 
As can be seen from the normalizations $N_i$ in Tab.~\ref{tab:fitpar}, 
we find that the dominant contribution to $D^{*+}$ mesons stems from valence charm quarks, $N_c=0.179$, as is expected. 
The total bottom FF, $D_{b+\bar{b}}^{D^{*+}}$ contributes much less, and only a very small, though important, fraction 
of the gluon momentum is used to produce $D^{*+}$ mesons. See the discussion of $pp$ data below.

%
%
\begin{table}[t!]
\caption{\label{tab:fitpar} Optimum parameters for our NLO FFs $D_i^{D^{*+}}(z,\mu_0)$
for positively charged $D^{*+}$ mesons
in the $\overline{\text{MS}}$ scheme at the input scale $\mu_0=m_c=1.3\,\mathrm{GeV}$; cf.\ Eq.~\eqref{eq:Dparam}. 
The bottom FF refers to $\mu_0=m_b=4.75\,\mathrm{GeV}$
and $\beta_g=10$ was kept fixed, see text.}
\begin{ruledtabular}
\begin{tabular}{lccc}
\label{tab:fitres}
flavor $i$ & $N_i$ & $\alpha_i$ & $\beta_i$\\ \hline
$c$ & 0.179 & 7.286 & 2.495\\
$b+\bar{b}$ & 0.084 &   3.654 &  6.832 \\
$g$ & 0.002 &   16.269 &   10\\
\end{tabular}
\end{ruledtabular}
\end{table}
%

In Tab.~\ref{tab:datasets} we list the data sets that pass the selection cuts on $z$ and $p_T$ as 
described in Sec.~\ref{subsec:data} above and are thus included in our fit. 
We show the number of data points that are fitted for each set along with the obtained individual $\chi^2$ values. 
In addition, we present the analytically obtained optimum normalization shifts $\mathcal{N}_i$ for each data set $i$.
They contribute to the total $\chi^2$ as specified in Eq.~\eqref{eq:chi2} and
according to the quoted experimental normalization uncertainties $\Delta \mathcal{N}_i$;
an entry $\mathcal{N}_i=1$ in Tab.~\ref{tab:datasets} indicates that 
normalization uncertainties are not provided by the experiment.
As can be seen, 96 data points from 3 different types of processes, 
SIA, single-inclusive hadron production, and in-jet fragmentation in $pp$ collisions are included in our
global QCD analysis of FFs for $D^{*+}$ mesons, 
yielding a $\chi^2$ per degree of freedom of 1.17 for our best fit.
%
\begin{table}[th!]
\caption{\label{tab:datasets} Data sets included in our global analysis, the corresponding 
optimum normalization shifts $\mathcal{N}_i$, and the individual $\chi^2$ including
the $\chi^2$ penalty from the determination of the normalization shift if applicable.}
\begin{ruledtabular}
\begin{tabular}{lccccc}
\label{tab:fitres}
                   & &  data   &           & \#data    &            \\
experiment & & type & $\mathcal{N}_i$ &  in fit & $\chi^2$\\ \hline
ALEPH \cite{Barate:1999bg} & & incl. & 0.991 & 17 & 31.0\\
OPAL \cite{Akers:1994jc}    &  & incl. &   1.000 & 9 & 6.5 \\
                  &  &$c$ tag &  1.002 &   9 & 8.6 \\
                  &  &$b$ tag & 1.002 & 9   & 5.6 \\ \hline
ATLAS \cite{Aad:2015zix} & & $D^{*\pm}$ & 1 & 5 & 13.8 \\
ALICE \cite{ALICE:2011aa} & $\sqrt{S} = 7$ TeV & $D^{*+}$ & 1.011 & 3 & 2.4\\
ALICE \cite{Abelev:2012vra}  & $\sqrt{S} = 2.76$ TeV &$D^{*+}$& 1.000 & 1 & 0.3\\
CDF \cite{Acosta:2003ax} & & $D^{*+}$ & 1.017 & 2 & 1.1 \\
LHCb \cite{Aaij:2015bpa} & $2 \le \eta \le 2.5$ & $D^{*\pm}$ & 1 & 5 & 8.2 \\
                    & $2.5 \le \eta \le 3$ & $D^{*\pm}$ & 1 & 5 & 1.6 \\
                    & $3 \le \eta \le 3.5$ & $D^{*\pm}$ & 1 & 5 & 6.5  \\
                    & $3.5 \le \eta \le 4$ & $D^{*\pm}$ & 1 & 1 & 2.8 \\[2mm] \hline
ATLAS \cite{Aad:2011td} & $25 \le \frac{p_T^\jet}{\text{GeV}} \le  30$ &$(\jet\,D^{*\pm})$ & 1 & 5 & 5.5\\[2mm]
                      & $30 \le \frac{p_T^\jet}{\text{GeV}} \le  40$ &$(\jet\,D^{*\pm})$ & 1 & 5 & 4.1\\[2mm]
                      & $40 \le \frac{p_T^\jet}{\text{GeV}} \le  50$ &$(\jet\,D^{*\pm})$ & 1 & 5 & 2.4\\[2mm]
                      & $50 \le \frac{p_T^\jet}{\text{GeV}} \le  60$ &$(\jet\,D^{*\pm})$ & 1 & 5 & 0.9\\[2mm]
                      & $60 \le \frac{p_T^\jet}{\text{GeV}} \le  70$ &$(\jet\,D^{*\pm})$ & 1 & 5 & 1.6\\[1mm] \hline
& & & & & \\
{\bf TOTAL:} & & & & 96 & 102.9
\end{tabular}
\end{ruledtabular}
\end{table}
%

\begin{figure}[t!]
\begin{center}
\includegraphics[width=0.5\textwidth]{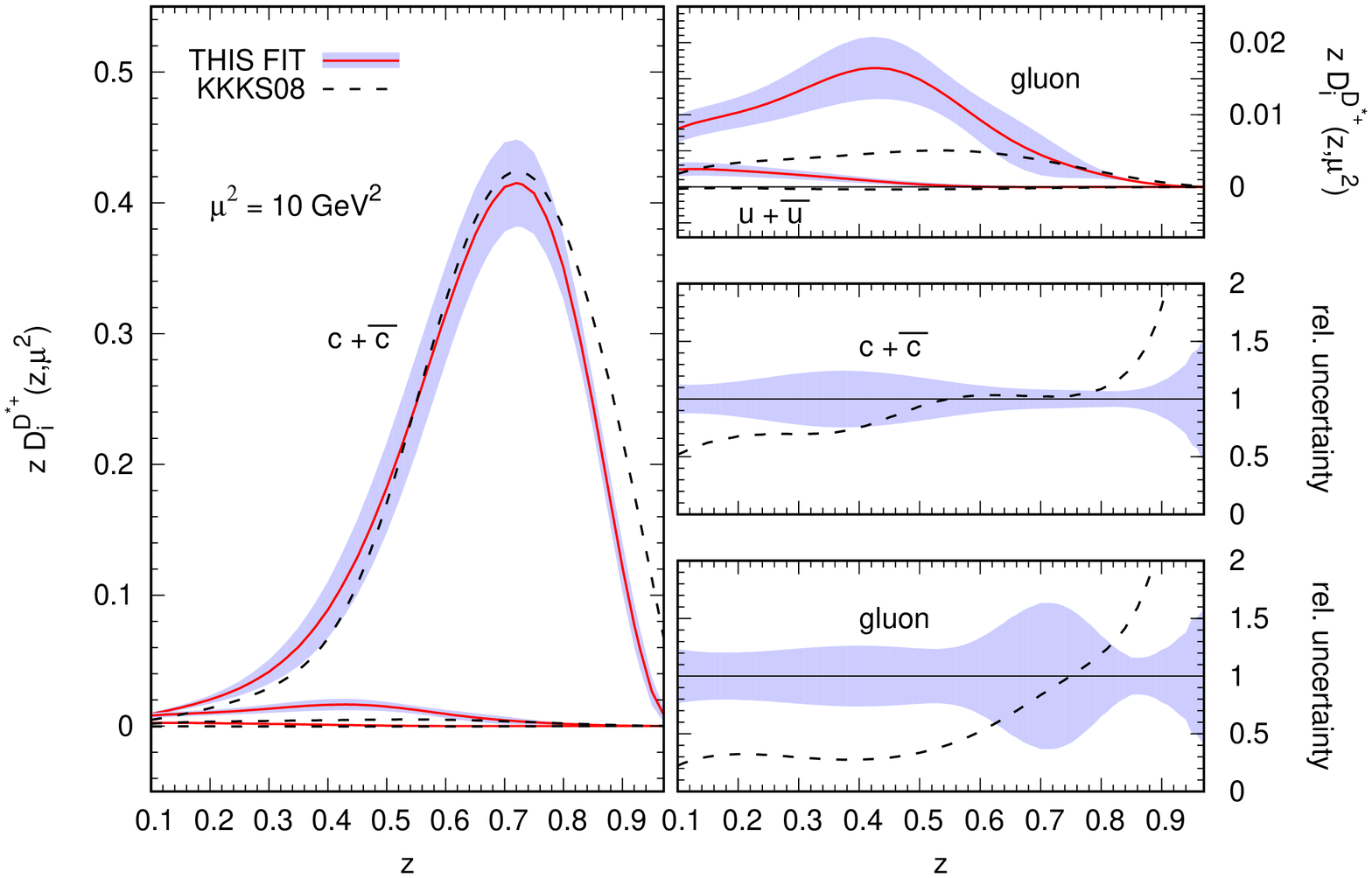}
\end{center}
\vspace*{-0.4cm}
\caption{Left-hand-side: our FFs $z D_i^{D^{*+}}(z,\mu^2)$ 
at scale $\mu^2 = 10\,\mathrm{GeV}^2$ (solid lines) along with the 
obtained uncertainty estimates (shaded bands). 
The dashed lines refer to the results of KKKS08 fit \cite{Kneesch:2007ey}.
Right-hand-side: to make the small gluon and $u+\bar{u}$ FFs better visible, they are
shown again in the upper panel.  
The middle and lower panels give the ratios of our uncertainty estimates (shaded bands) and
the KKKS08 fit relative to our best fit
for the $c+\bar{c}$ and the gluon FF, respectively.
\label{fig:FFs_at_10}
}
%
\begin{center}
\includegraphics[width=0.5\textwidth]{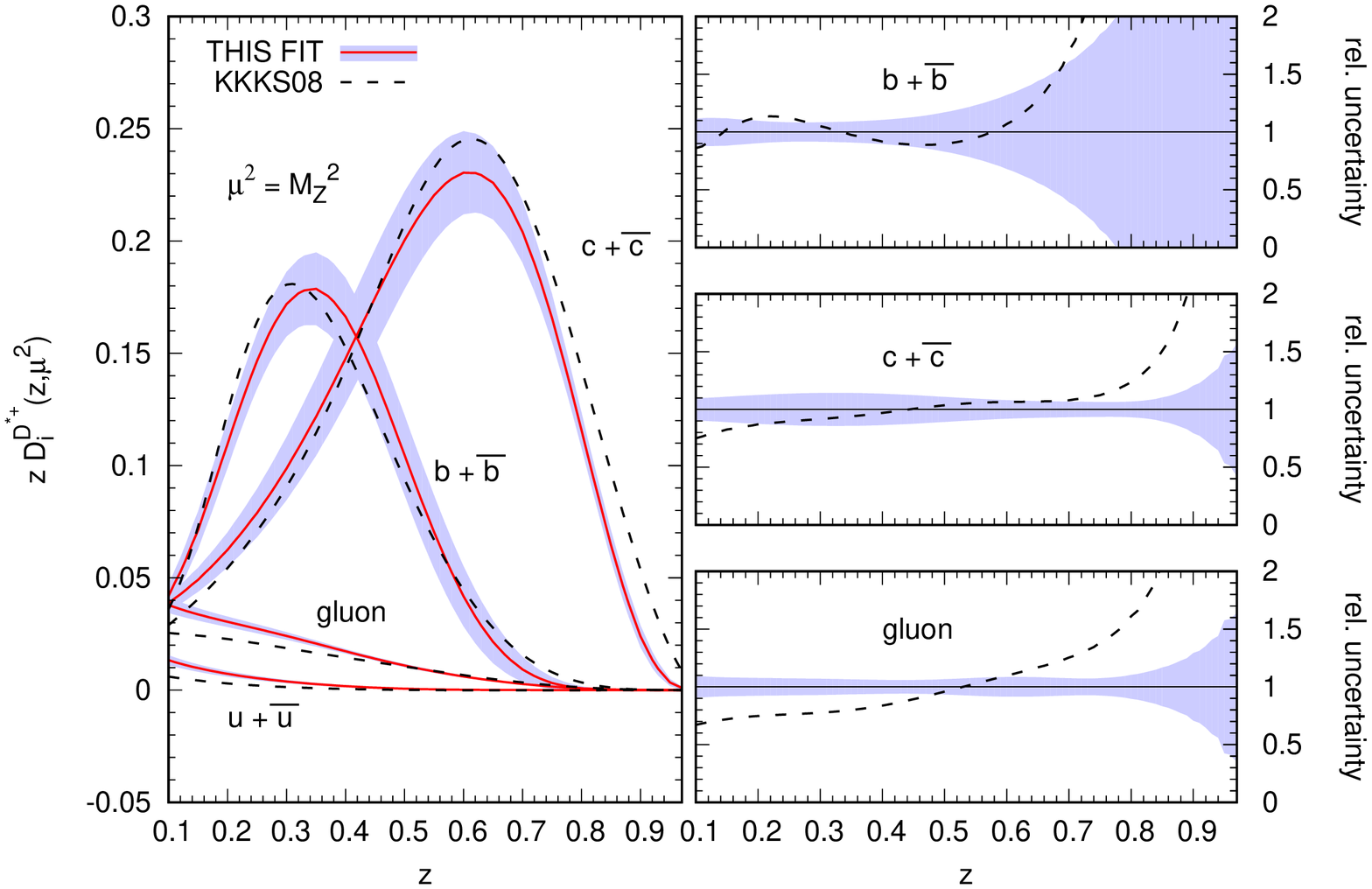}
\end{center}
\vspace*{-0.4cm}
\caption{Similar to Fig.~\ref{fig:FFs_at_10} but now for $\mu^2 = M_Z^2$ and including the total bottom FF.
Here, the upper right panel shows the relative uncertainties and comparison to KKKS08 for the total $b+\bar b$ FF.
\label{fig:FFs_at_LEP}
}
\end{figure}

The so obtained FFs are shown for two representative scales $\mu^2 = 10\,\mathrm{GeV}^2$ and $\mu^2 = M_Z^2$
in Fig.~\ref{fig:FFs_at_10} and \ref{fig:FFs_at_LEP}, respectively, along with our 
uncertainty estimates (shaded bands) based on the Hessian method with $\Delta \chi^2=4$ \cite{ref:sets}, 
see Sec.~\ref{subsec:errors}.
%
As the gluon and the unfavored light quark contributions turn out to be very small compared to the dominant 
charm-to-$D^{*+}$ FF, we show them again for the sake of better legibility in the top right panel of Fig.~\ref{fig:FFs_at_10}.
Notice that we just show the total $u + \bar{u}$ FF as one example of the unfavored light quark and $\bar{c}$ FFs, which
are all the same as they are generated solely by QCD evolution from a vanishing input distribution, see
Sec.~\ref{subsec:parametrization}. This affects also the uncertainty estimates for these FFs which arise, again,
just from evolution, i.e., mainly by propagating the uncertainties of the gluon FF. Hence, there is no direct access
to the uncertainties of the unfavored light quark and $\bar{c}$ FFs, such that they have to be taken with a grain of salt.
Since none of the presently available data sets is sensitive to the unfavored FFs into a $D^{*+}$ meson,
in contrast to the also small gluon FF, one is forced to make some assumption about them. 
In any case, light quarks are expected to fragment mainly into 
light mesons such as pions and kaons, and their contribution to $D^{*}$ meson production should be small. 
Therefore, our choice of a vanishing input distribution for all unfavored FFs appears to be reasonable.
We note that a similar assumption was made in the KKKS08 fit \cite{Kneesch:2007ey}.

It is instructive to compare the results of our FFs into $D^{*+}$ mesons to those obtained in
the KKKS08 fit \cite{Kneesch:2007ey} that is based only on SIA data and includes the by now
obsolete and withdrawn BELLE data. The KKKS08 results for $z D_i^{D^{*+}}(z,\mu^2)$ and the
ratio to our FFs are shown as dashed lines in
Figs.~\ref{fig:FFs_at_10} and \ref{fig:FFs_at_LEP}.
As can be seen, one of the main differences is that our fit returns a significantly larger gluon contribution 
compared to KKKS08 at intermediate values of $z$ which might be related to the fact that also the
gluon FF starts from a vanishing input in the KKKS08 fit. However, both the inclusive high-$p_T$ and,
in particular, the in-jet fragmentation data, for the first time included in our global analysis,
demand a non-zero gluon FF at our input scale in order to arrive at a satisfactory
description of the data; see also the detailed comparisons to the inclusive and in-jet $pp$ data below.
One also notices, that the two valence charm FFs are somewhat shifted in $z$ with respect to each other
and that also the height of the peak is different. This is most likely caused by the different
sets of SIA data included in our and the KKKS08 analyses. Also, the KKKS08 fit does not include any uncertainty estimates.

Finally, in Fig.~\ref{fig:FFs_at_LEP}, for $\mu=M_Z$, we also show the bottom-to-$D^{*+}$ FF which
starts to evolve from a non-zero input above the threshold $\mu_0=m_b$, see Sec.~\ref{subsec:parametrization}.
The total $b+\bar{b}$ FF turns out to be quite similar to the one obtained in the KKKS08 analysis.
This is to be expected as the bottom FF is largely constrained by the bottom-tagged data of the OPAL
collaboration which are included in both fits.

\subsection{Detailed comparison to data}
In this section we compare theoretical calculations based on the results of 
our global QCD analysis with the available data.  
Throughout, we shall also show uncertainty bands obtained with the Hessian
sets for $\Delta \chi^2=4$ as discussed in Sec.~\ref{subsec:errors}.
In addition, we perform all calculations with the FFs provided by
Ref.~\cite{Kneesch:2007ey}. 
Notice that \cite{Kneesch:2007ey} provides two sets of FFs which differ in the way
they include finite charm quark and $D^*$ meson mass effects. Since we work in the ZMVFNS,
we choose, as in Figs.~\ref{fig:FFs_at_10} and \ref{fig:FFs_at_LEP} above,
the corresponding KKKS08 set of FFs without quark mass effects in order
to arrive at a meaningful comparison with our results.
However, according to \cite{Kneesch:2007ey}, the KKKS analysis, some kinematic corrections 
due to the mass of the $D^{*}$ meson have been retained in all their fits 
beyond the standard theoretical framework based on factorization, 
which might be the source of some of the differences we observe at small $z$.

\begin{figure}[th!]
\begin{center}
\includegraphics[width=0.5\textwidth]{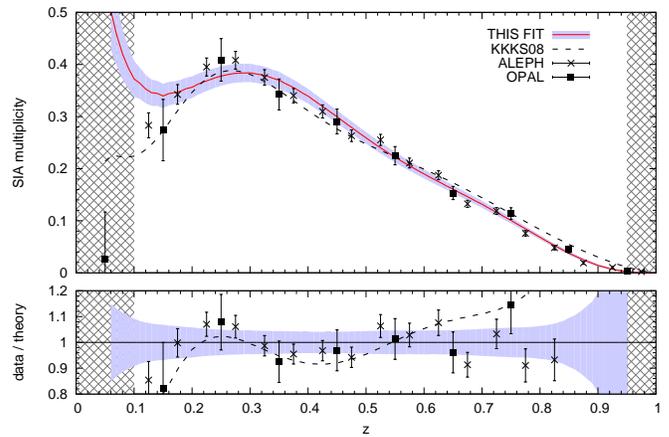}
\end{center}
\caption{The SIA multiplicity data from LEP \cite{Barate:1999bg, Akers:1994jc} at $Q=M_Z$
are shown together with theory calculations using our best fit (solid lines) and FFs of KKKS08 (dashed lines).
The shaded bands refer to our uncertainty estimates and the hatched areas 
are excluded from the fit, see text.
\label{fig:SIA_LEP}
}
\end{figure}

We start with a study of the inclusive SIA data with identified $D^{*\pm}$ mesons.
The upper panel of Fig.~\ref{fig:SIA_LEP} shows the LEP data at $Q=M_Z$ from 
the ALEPH \cite{Barate:1999bg} and OPAL \cite{Akers:1994jc} collaboration along with
the theory calculations for the SIA multiplicities at NLO accuracy 
as defined in Eq.~\eqref{eq:TL}. The solid and dashed lines are obtained with our
best fit, labeled as ``this fit'' throughout this section, and the FFs of KKKS08.
The ratio of data over theory, our relative uncertainty
estimates, and the ratio between KKKS08 and our best fit are given in the lower panel of
Fig.~\ref{fig:SIA_LEP}.
The hatched regions with $z < z_{\min}=0.1$ and $z > 0.95$ are excluded from our
fit as discussed in Sec.~\ref{subsec:data}. The latter cut, which has no impact on
the fit and in total only removes a single data point from our analysis, 
is imposed due to the presence of potentially large logarithms as $z\to 1$,
which cannot be properly accounted for in a fixed-order calculation.
%
%

As can be already anticipated from the individual $\chi^2$ values listed in Tab.~\ref{tab:fitres},
we find that our fit describes the inclusive SIA data very well, and our Hessian
uncertainty estimates reflect the experimental uncertainties except for the 
data points with the lowest value of $z$ in each data set.
Both sets of FFs describe the data equally well in the intermediate $z$-region.
Towards larger values of $z$, the KKKS08 FFs overshoot the LEP SIA data significantly,
which might be related to some tension with the CLEO and the by now withdrawn BELLE data,
that are both included in their fit.
In the small-$z$ region around our cut $z_{\min}$, the KKKS08 fit agrees slightly better with the data
which might indicate some signs of a breakdown of the massless framework which we pursue in our
analysis. In addition, we note, that the fixed-order evolution of FFs becomes more and more unstable 
towards smaller values of $z$, see e.g. \cite{Anderle:2016czy}. 
Eventually, this can result in unphysical negative values for the FFs and the cross sections. 
The onset of this pathological behavior depends of the fit parameters and might, in part, also
be responsible for the KKKS08 results to start to drop. 
At smaller values of $z$ than shown in Fig.~\ref{fig:SIA_LEP}, well below our cut $z_{\min}$, 
even our, still rising SIA multiplicity will
start to drop and eventually reach unphysical, negative values.

\begin{figure}[th!]
\begin{center}
\includegraphics[width=0.5\textwidth]{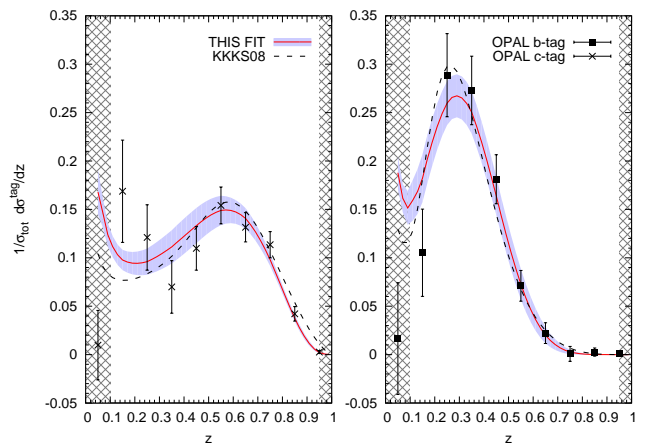}
\end{center}
\caption{Charm (left) and bottom (right panel) tagged SIA multiplicities 
for charged $D^{*\pm}$ mesons from OPAL \cite{Akers:1994jc} at $Q=M_Z$ compared with theory calculations
at NLO accuracy using our best fit (solid lines) and the KKKS08 FFs (dashed lines).
The shaded bands refer to our uncertainty estimates based on the Hessian method.
The hatched areas are excluded from our fit.
\label{fig:SIA_LEP_flavortag}
}
\end{figure}

In Fig.~\ref{fig:SIA_LEP_flavortag}, we show the charm and bottom flavor-tagged data 
from OPAL \cite{Akers:1994jc} which are normalized to the total hadronic cross section.
The bottom-tagged data are particularly instrumental in constraining the 
total bottom-to-$D^{*+}$ FF in both our and the KKKS08 fit.
As can be seen, both theoretical results describe the flavor-tagged data equally well, which have rather
large uncertainties compared to the inclusive results shown in Fig.~\ref{fig:SIA_LEP}.
%
%

Following the order of processes as discussed in Sec.~\ref{sec:tech}, we next
consider the single-inclusive, high-$p_T$ production of $D^{*}$ mesons in hadronic collisions. 
Since we are working in the ZMVFNS, we are especially interested in data where the observed $D^{*}$ meson has
a transverse momentum $p_T$ much larger than the charm quark or the $D^*$ meson mass, i.e.,
$p_T \gg m_{D^{*}} \sim m_c \approx 2\,\mathrm{GeV}$.
As discussed in Sec.~\ref{subsec:data}, we employ a rather stringent cut of $p_T > 10\,\mathrm{GeV}$
in our global analysis. However, we will demonstrate that the so obtained FFs work unexpectedly well in 
describing single-inclusive $D^*$ meson cross sections down to much smaller values of
$p_T$ around $5\,\mathrm{GeV}$.

\begin{figure}[t]
\begin{center}
\includegraphics[width=0.5\textwidth]{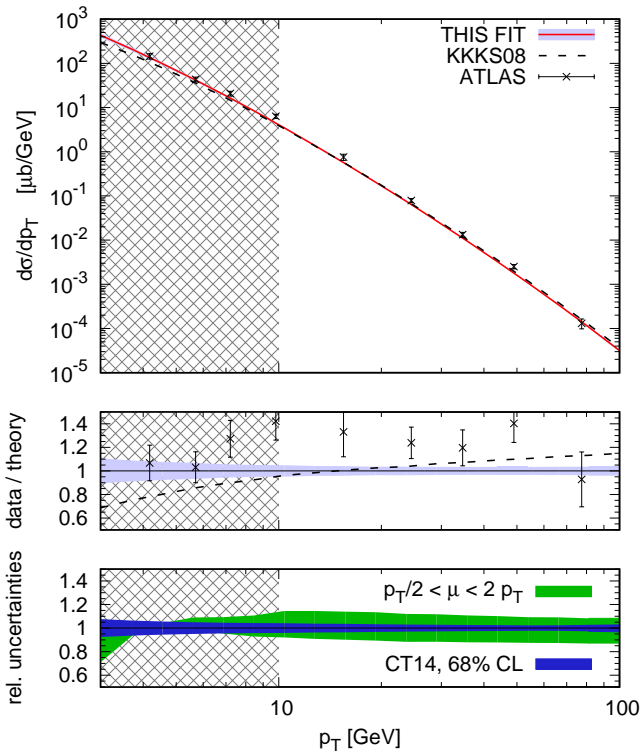}
\end{center}
\caption{Upper panel: our NLO result (solid line) for the single-inclusive, high-$p_T$ cross section for $D^{*\pm}$ 
meson production in $pp$ collisions at $\sqrt{S}=7\,\mathrm{TeV}$ and integrated over rapidity
$|\eta|<2.1$ compared to data from the ATLAS collaboration~\cite{Aad:2015zix}
and a calculation using the KKKS08 FFs (dashed line). The middle panel shows the corresponding ratios
to our result. The shaded bands refer to our uncertainty estimates based on the Hessian method.
The lower panel illustrates relative theoretical uncertainties due to variations of the 
scale $\mu$ in Eq.~\eqref{eq:xsec_pphx} (outer shaded band) and the error estimate of the
CT14 PDFs (inner shaded band) which we have rescaled to $68\%$ C.L., see text.
The hatched areas are excluded due to the cut $p_T>10\,\mathrm{GeV}$ imposed on our fit. 
\label{fig:pph_ATLAS}
}
\end{figure}
%
In this respect, the most relevant data set is the one presented by the ATLAS collaboration \cite{Aad:2015zix}, 
shown in Fig.~\ref{fig:pph_ATLAS},
which covers the range $3.5\,\mathrm{GeV}< p_T < 100\,\mathrm{GeV}$ at a $pp$ c.m.s.\ energy of $\sqrt{S}=7\,\mathrm{TeV}$
integrated over the mid rapidity range $|\eta|<2.1$.
In the upper panel, a comparison of the ATLAS data with calculations at NLO accuracy is presented 
based on our best fit and KKKS08 $D^{*\pm}$ FFs and using Eq.~\eqref{eq:xsec_pphx}; again, data in the hatched area, 
i.e., below  our cut $p_T^{\min}=10\,\mathrm{GeV}$, data are not included in our global analysis.
The middle panel gives the ratios of the KKKS08 prediction and the ATLAS data with respect to our NLO calculation.
In addition, it illustrates the uncertainty estimates (shaded bands) obtained from our Hessian sets of $D^{*\pm}$ FFs.

Both sets of FFs provide a satisfactory description of the data at NLO accuracy in the ZMVFNS
even well below $p_T=10\,\mathrm{GeV}$. About $50\%$ of the $D^{*\pm}$ mesons at $p_T\simeq 10\,\mathrm{GeV}$ 
originate from gluon fragmentation which drops down to approximately $40\%$ at the highest $p_T$ measured
by ATLAS. In view of the sizable differences between our and the KKKS08 gluon FF illustrated 
in Figs.~\ref{fig:FFs_at_10} and \ref{fig:FFs_at_LEP}, the similarity of the NLO cross sections is a remarkable result 
and indicates that the inclusive $p_T$ spectra only constrain certain $z$-moments of the gluon FF rather than its detailed $z$ shape. 
The differences between the two sets of FFs in Figs.~\ref{fig:FFs_at_10} and \ref{fig:FFs_at_LEP},
in particular, the gluon FF, will be much more pronounced when we turn to the in-jet fragmentation data below.

In the lower panel of Fig.~\ref{fig:pph_ATLAS}, we show other important sources of
theoretical uncertainties associated with a pQCD calculation of $pp$ cross sections based on, e.g., 
Eq.~\eqref{eq:xsec_pphx}. The outer shaded bands illustrates the ambiguities due to 
simultaneous variations of the factorization and renormalization scales in the range 
$p_T / 2 < \mu < 2 p_T$. As can be seen, in the $p_T$ range used in our fit, this results in a roughly constant
relative uncertainty of about $10\%$. The theoretical error related to PDF uncertainties are estimated
with the Hessian sets provided by the CT14 collaboration \cite{Dulat:2015mca}. They turn out to be
smaller than QCD scale uncertainties and are at a level of about $5\%$ as can be inferred 
from the  inner shaded bands in the lower panel of Fig.~\ref{fig:pph_ATLAS}.
To be compatible with our estimates of the one-sigma uncertainties of the $D^{*\pm}$ FFs we follow
Ref.~\cite{Watt:2011kp} and rescale the available CT14 Hessian sets from the $90\%$
to the $68\%$ confidence level by a applying constant factor $1/1.645$.
 
\begin{figure}[th!]
\begin{center}
\includegraphics[width=0.5\textwidth]{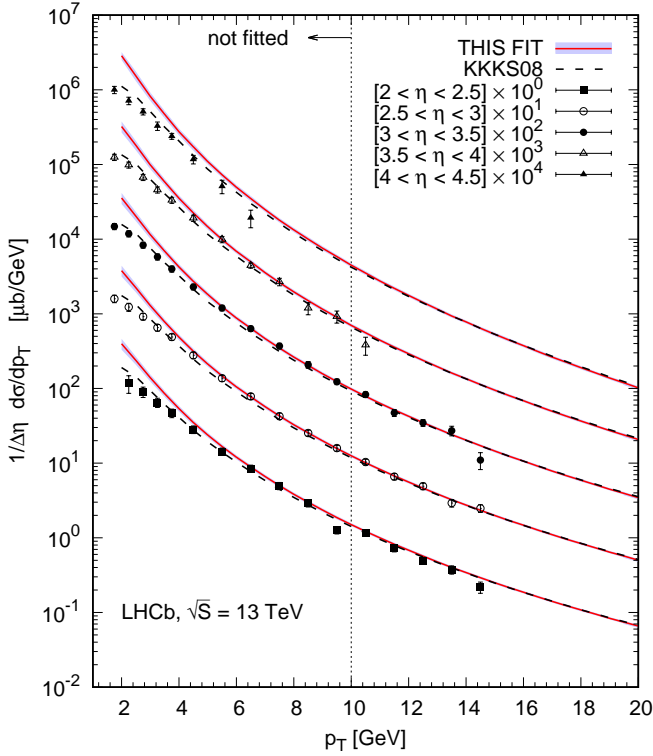}
\end{center}
\caption{The $p_T$-spectra of $D^{*\pm}$ mesons at $\sqrt{S} = 13$ TeV for five different bins in rapidity,
normalized to the width $\Delta \eta$ of each rapidity bin, as measured by LHCb \cite{Aaij:2015bpa}.
To better delineate the data, each rapidity bin was multiplied with an increasing power of 10.
The NLO calculations using our best fit and the KKKS08 FFs are shown as solid and dashed lines,
respectively. As before, the shaded bands refer to our  uncertainty estimates, and data
below $p_T^{\min}=10\,\mathrm{GeV}$ are excluded from out fit.
\label{fig:pph_LHCb13}
}
\end{figure}
\begin{figure}[th!]
\begin{center}
\includegraphics[width=0.5\textwidth]{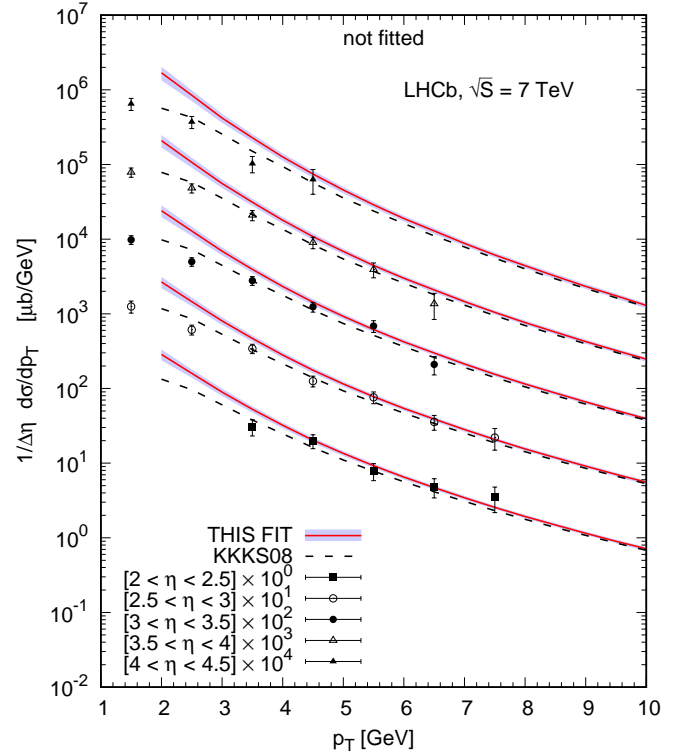}
\end{center}
\caption{Same as Fig.~\ref{fig:pph_LHCb13} but now for $\sqrt{S} = 7$ TeV. 
Note, that all data points are below our cut on $p_T$.
\label{fig:pph_LHCb7}
}
\end{figure}

%
A large amount of data points for inclusive $D^{*\pm}$-meson production have been presented by the LHCb collaboration. 
They measured the single-inclusive $D^{*\pm}$ production cross section 
at forward rapidities $\eta$ for two different 
c.m.s.\ energies, $\sqrt{S} = 7$~TeV~\cite{Aaij:2013mga} and $13\,\mathrm{TeV}$~\cite{Aaij:2015bpa}. 
For each c.m.s.\ energy, the data are presented in five bins of rapidity in the range 
from $\eta = 2$ up to $\eta = 4.5$. Compared to the mid rapidity data shown in Fig.~\ref{fig:pph_ATLAS},
the LHCb data are limited to smaller values of $p_T$. Nevertheless, several data points from the
$\sqrt{S} = 13\,\mathrm{TeV}$ run \cite{Aaij:2015bpa} are above our cut $p_T^{\min}=10\,\mathrm{GeV}$ except for the most
forward rapidity bin $4<\eta<4.5$ but, unfortunately, none of the data points taken at
$\sqrt{S} = 7\,\mathrm{TeV}$~\cite{Aaij:2013mga} passes the cut. 
Both sets of data are shown in Figs.~\ref{fig:pph_LHCb13} and \ref{fig:pph_LHCb7} and compared to the
results of NLO calculations based on our and the KKKS08 set of FFs.
We note that the more forward the rapidity interval, the more important is the role of 
gluon fragmentation in producing the observed $D^{*\pm}$ mesons,
a feature that has already been observed for the production of lighter hadrons at the LHC \cite{Sassot:2010bh}.
For instance, at $\sqrt{S}=7\,\mathrm{TeV}$ around $80\%$ of the $D^{*\pm}$ mesons at $p_T\simeq 5\,\mathrm{GeV}$
originate from gluons. Since forward data also sample on average larger values of $z$ \cite{Sassot:2010bh},
the LHCb data nicely complement the mid rapidity data by ATLAS.

%
As for the ATLAS data, both sets of FFs also give an equally good description of the LHCb data shown in
Fig.~\ref{fig:pph_LHCb13} for $p_T>p_T^{\min}$, as can be also inferred from Tab.~\ref{tab:fitres},
and they continue to follow the data well below our cut, down to about $5\,\mathrm{GeV}$. Also
the data taken at $\sqrt{S}=7\,\mathrm{TeV}$, that are not included in our fit, are well described
down to $p_T\simeq 5\,\mathrm{GeV}$ except for the most forward bin $4<\eta<4.5$.
The KKKS08 FFs follow the trend of the data even further down to the lowest $p_T$ values shown in 
Figs.~\ref{fig:pph_LHCb13} and \ref{fig:pph_LHCb7}; for the sake of applicability of pQCD, we refrain
from showing comparisons to the LHCb data below $p_T=2\,\mathrm{GeV}$. This feature of the KKKS08 fit, which is
unexpected in a ZMVFNS approach, might be due to the inclusion of finite hadron mass corrections
in their fit of SIA data, that are, however, beyond the factorized framework outlined in Sec.~\ref{sec:tech}
and adopted by us.
It is also interesting to notice that there are some indications for a mild tension
between the ATLAS and the LHCb data in our global fit. The ATLAS data alone would prefer a somewhat
larger gluon-to-$D^{*+}$ meson FF as can be inferred from the middle panel of Fig.~\ref{fig:pph_ATLAS}. 
This would yield a significantly better fit of the ATLAS data in terms of $\chi^2$ even when the
in-jet fragmentation data, which we shall discuss next, are included in the fit. The latest, revised version 
of the LHCb data \cite{Aaij:2015bpa} does not tolerate, however, such an increased gluon FF in our global analysis.

\begin{center}
\begin{figure*}[th!]
\begin{center}
\includegraphics[width=0.8\textwidth]{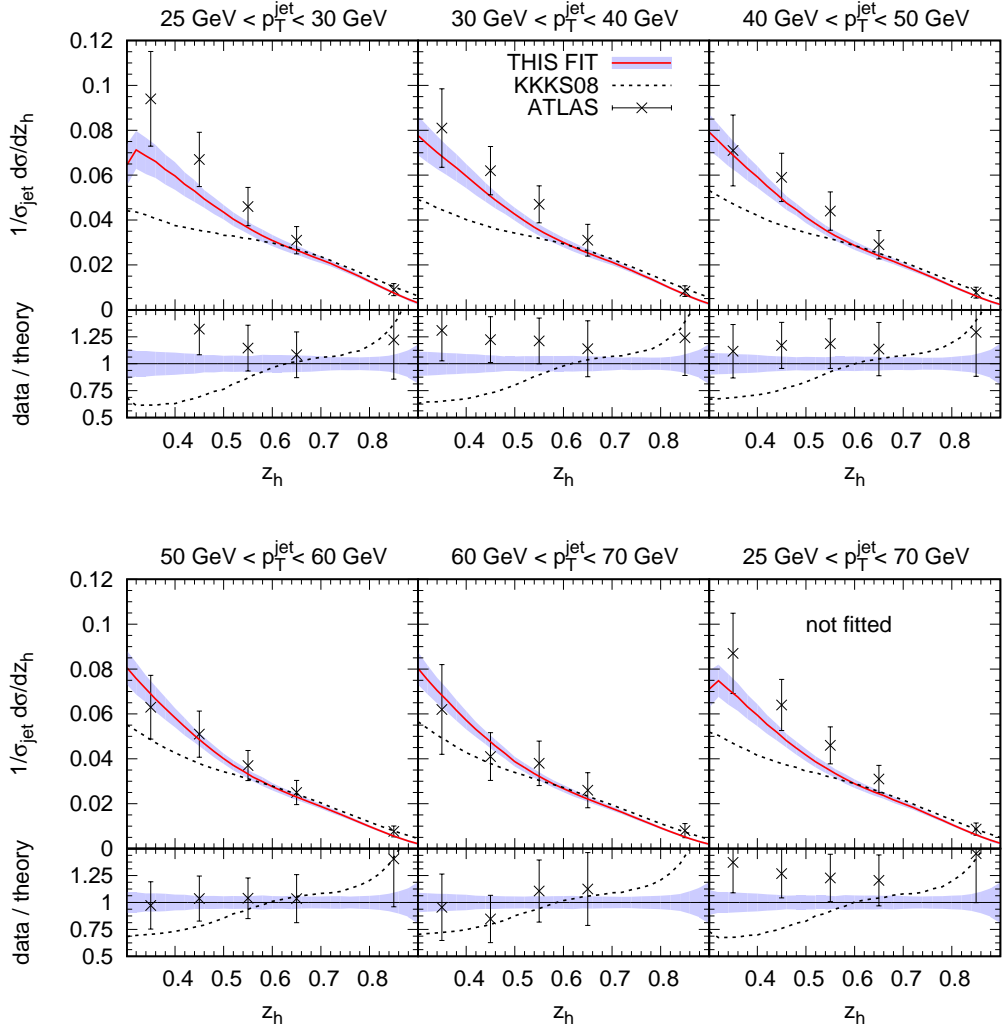}
\end{center}
\caption{Data on in-jet-fragmentation into $D^{*\pm}$ mesons measured at $\sqrt{S}=7$ TeV as a function of the
momentum fraction $z_h$ in five bins of $p_T^\jet$ integrated over rapidity $|\eta^\jet|<2.5$
as provided by ATLAS \cite{Aad:2011td}. The combination of all $p_T^\jet$ bins (lower right)
is only shown for comparison and is not included in our fit to avoid double-counting. 
In each panel, NLO results
obtained with our best fit (solid lines) and the KKKS08 (dashed lines) FFs are shown.
The shaded bands refer to uncertainty estimates based on our Hessian uncertainty sets.
In the lower panels of each plot, the ratio of the data and the KKKS08 prediction
with respect to our NLO result are given.
\label{fig:jet_h}
}
\end{figure*}
\end{center}
%
We refrain from showing comparisons of our theoretical results with the ALICE and CDF data on single-inclusive,
high-$p_T$ $D^{*+}$ meson production. As can be seen from Tab.~\ref{tab:datasets}, the few data points which
pass our cut on $p_T$ are very well reproduced by our fit. Again, adopting the KKKS08 set of FFs leads to
a similar description of these data, assuming $D_i^{D^{*+}} = D_i^{D^{*\pm}}/2$.

Finally, we turn to data on in-jet production, which, in this paper, are considered
for the first time in a global QCD analysis of FFs and, hence, represent the centerpiece
of our phenomenological studies.  
The relevant QCD formalism to compute in-jet production in the standard factorized framework
at NLO accuracy was sketched in Sec.~\ref{sec:jethX}. The main and novel asset of this process, 
as compared to single-inclusive hadron production in $pp$ collisions, is the fact that in-jet 
data probe the parton-to-hadron FFs {\em locally} in the momentum fraction $z$ in the LO approximation. 
Therefore, one anticipates a much improved sensitivity to the, in particular, $z$-dependence of the
gluon FF also beyond LO accuracy than from single-inclusive probes. In the latter case,
we have just found that two rather different gluon FFs, ours and the one from the KKKS08 fit,
can result both in a good description of the existing data, cf.\ Figs.~\ref{fig:pph_ATLAS} and
\ref{fig:pph_LHCb13} above.

Specifically, for the in-jet production of $D^{*\pm}$ mesons, it was found in Ref.~\cite{Chien:2015ctp} 
that the cross section computed with the KKKS08 set of FFs falls significantly short of 
the corresponding yields observed by ATLAS \cite{Aad:2011td}. 
The authors of Ref.~\cite{Chien:2015ctp} observed that by increasing the
KKKS08 gluon FFs ad hoc by a $z$-independent factor of 2 would help to better describe the ATLAS data. 
However, such a modified gluon FF would then significantly overshoot the 
single-inclusive $pp$ data for $D^{*}$-mesons. Clearly, to address this issue reliably and in detail, a simultaneous
global QCD analysis of all relevant probes comprising SIA, and single-inclusive and in-jet production in
proton-proton collisions is absolutely essential.

From Fig.~\ref{fig:jet_h} and Tab.~\ref{tab:fitres} one can gather that our global fit yields an
excellent description of the in-jet data by ATLAS in all five bins of the jet's transverse momentum
without compromising the comparison to SIA or single-inclusive $pp$ data.
A corresponding calculation with the KKKS08 set of FFs falls short of the data for 
momentum fractions $z_h\lesssim 0.6$ of the $D^{*\pm}$ meson,
as was already observed in Ref.~\cite{Chien:2015ctp}.
In fact, the $z$-dependence of the NLO calculations with the two different sets of FFs
very closely follow the corresponding dependence
of the gluon-to-$D^{*\pm}$ FF illustrated for two different scales in 
Figs.~\ref{fig:FFs_at_10} and \ref{fig:FFs_at_LEP}.
The main difference between our analysis and the KKKS08 extraction of FFs is that 
we allow for a non-zero gluon FF at our initial scale which appears to be necessary in order to 
achieve a good global fit of all data; recall that the KKKS08 analysis was based only on SIA data
where some assumption about the gluon FF has to be made. The quark FFs, in particular, the 
charm FF, adjust accordingly in the fit but play only a very minor role in computations
of $pp$ cross sections in the $p_T$ range currently covered by experiment.
Finally, we note that theoretical uncertainties due to the choice of scale $\mu$ and from
ambiguities in the adopted set of PDFs are of similar size as we have estimated for the single-inclusive data;
cf.\ the lower panel of Fig.~\ref{fig:pph_ATLAS} and Ref.~\cite{Kaufmann:2015hma}.

Our case study of $D^{*\pm}$ clearly reveals how powerful in-jet data can be in further constraining FFs. 
Based on the framework developed and applied in this paper, in-jet data can be straightforwardly included in
any future global fit of FFs once such data become available.

\section{Conclusions and Outlook \label{sec:conclusions}}
%
We have presented the first global QCD analysis of fragmentation functions that makes use of
in-jet data besides the usual sets of experimental results stemming from single-inclusive hadron production 
in electron-positron annihilation and proton-(anti)proton collisions.
The necessary technical framework to incorporate in-jet fragmentation 
data consistently into a global fit at next-to-leading order accuracy 
was outlined in detail, and an implementation within the Mellin moment technique was 
given and henceforth adopted in all our phenomenological studies.

As a case study, we have analyzed available data for charged $D^*$ mesons in terms
of parton-to-$D^{*+}$ meson fragmentation functions. An excellent global
description of all the different processes included in the fit was achieved. In particular, the
in-jet fragmentation data have been shown to be of great importance in pinning down
the otherwise largely unconstrained momentum fraction dependence of the 
gluon fragmentation function. Compared to the only other previously available set of 
$D^{*\pm}$ fragmentation function, that was based solely on electron-positron annihilation data,
we obtain a rather different momentum dependence for the hadronization of gluons in order to
describe the in-jet data.

In addition to our optimum fit, we have also, for the first time, estimated the uncertainties
of charged $D^*$ meson fragmentation functions. To this end, we have applied the Hessian method.
The obtained Hessian sets provide a straightforward way to propagate our estimated uncertainties
to any other process of interest. Apart from the experimental uncertainties that are incorporated in
the Hessian sets, we have illustrated the importance of other, theoretical sources of ambiguities
comprising the actual choice of renormalization and factorization scales and corresponding uncertainties
of parton distribution functions which are needed in calculations of any hadronic collision process.

For the time being, we have adopted the ZMVFNS throughout our global analysis, i.e., we have imposed
rather stringent cuts on the minimum transverse momentum of the $D^{*\pm}$ mesons for data to be included
in our fit. We have demonstrated, however, that our fit gives a reasonable description 
of single-inclusive data from the LHC both at mid and forward rapidities even down to significantly smaller
values of transverse momentum of about $5\,\mathrm{GeV}$.

We believe that in-jet data will prove very valuable in the future 
in any upcoming analysis of fragmentation functions, in particular, in further constraining the 
detailed momentum dependence of the hadronization of gluons.
The framework developed and applied in this paper can be straightforwardly generalized to
incorporate in-jet data in any future global fit of FFs once such data become available.
We plan to extend our phenomenological studies to charged and neutral $D$ mesons in the near
future. We note that the theoretical framework for the in-jet production of hadrons at
next-to-leading order accuracy has been recently extended to include also photons
\cite{Kaufmann:2016nux}. The fragmentation into photons is so far only rather poorly understood 
and constrained by data. Again, we expect 
any upcoming in-jet data to be very valuable in a new extraction of photon fragmentation functions. 
Finally, we plan to study in detail the impact of the resummation of logarithms of the jet size parameter $R$. 
By making use of the results for the in-jet fragmentation of hadrons derived within the SCET formalism, 
it is possible to extract fragmentation functions at a combined accuracy of NLO+NLL$_{\rm R}$.

\section*{Acknowledgments}
%
We are grateful to Y.-T.\ Chien, Z.-B.\ Kang, E.\ Mereghetti, N.\ Sato, and W.\ Vogelsang 
for helpful discussions and comments.
D.P.A.\ was supported by the Lancaster-Manchester-Sheffield Consortium for Fundamental Physics under STFC Grant No. ST/L000520/1.
T.K.\ was supported by
the Bundesministerium f\"{u}r Bildung und Forschung (BMBF) under 
grant no.\ 05P15VTCA1.
F.R.\ is supported by the LDRD Program of Lawrence Berkeley National Laboratory, 
the U.S.\ Department of Energy, Office of Science and Office of Nuclear Physics under 
contract number DE-AC02-05CH11231. I.V.\ is supported by the  US  Department  of  Energy, 
Office of Science under Contract No. DE-AC52-06NA25396 and the DOE Early Career Program.


%

\begin{thebibliography}{99}
\bibitem{ref:sets} A {\sc Fortran} code of our optimum fit of $D^{*+}$, $D^{*-}$ and $D^{*\pm}$ FFs and the corresponding
Hessian uncertainty sets are available upon request from the authors.
\end{thebibliography}
\end{document}